\documentstyle[prb,preprint,aps]{revtex}
\font\ee=msbm10 scaled \magstep1
\font\ee=msbm10 scaled \magstep1

\tightenlines
\date{\today}
\begin{document}
\thispagestyle{empty}
%\begin{flushright}
%\begin{tabular}{l}
%ANL-HEP-PR-00-119 
%\end{tabular}
%\end{flushright}
\vskip.5in minus.2in
\begin{center}
{\bf \Large The Moyal-Lie Theory of Phase Space Quantum Mechanics}

\vskip.5in minus.2in 

T. Hakio\u{g}lu \\
Physics Department, Bilkent University, 06533 Ankara, Turkey \\
and \\ 
Alex J. Dragt \\
Physics Department, University of Maryland, College Park MD 20742-4111, 
USA
\end{center}
\begin{abstract}
A Lie algebraic approach to the unitary transformations in Weyl 
quantization is discussed. This approach, being formally equivalent 
to the $\star$-quantization is an extension of the classical Poisson-Lie 
formalism which can be used as an efficient tool 
in the quantum phase space transformation theory.
\end{abstract}

\vskip0.3truecm
The purpose of this paper is to show that the Weyl correspondence in the
quantum phase space
can be presented in a quantum Lie algebraic perspective which in some
sense derives analogies from the Lie algebraic approach to classical
transformations and Hamiltonian vector fields. In the classical case 
transformations of functions $f(z)$ of the phase space $z=(p,q)$  
are generated by the generating functions ${\cal G}_{\mu}(z)$. 
If the generating functions are elements of a   
set, then this set usually contains a subset closed under the action the 
Poisson Bracket (PB), so defining an algebra. To be more  
precise one talks about different representations of the generators and  
of the phase space algebras. The second well known representation  
is the so called adjoint representation of the Poisson algebra of the 
${\cal G}_{\mu}$'s in terms of the classical 
Lie generators $L_{{\cal G}_{\mu}}$ over the Lie algebra. 
These are the Hamiltonian vector fields with the Hamiltonians  
${\cal G}_{\mu}$. By Liouville's theorem, they characterize  
an incompressible and covariant phase space flow. 

The quantum Lie approach can be based on a parallelism with the classical  
one above. In this case the 
unitary transformations are represented by a set of quantum generating 
functions ${\cal A}_\mu(z)$ which are the quantum partners of the 
${\cal G}_\mu$'s. The non commutativity  
is encoded in a new multiplication rule, i.e. $\star$-product in $z$. 
The quantum Lie generators $\hat{V}_{{\cal A}_\mu}$ over the Lie 
bracket are the adjoint 
representations of the generating functions ${\cal A}_\mu(z)$ over the 
Moyal bracket and they correspond to quantum counterparts of the classical
Hamiltonian vector fields $L_{{\cal G}_\mu}$. 
They are represented by an infinite number of phase space derivatives.  
The only exception is the generators of linear 
canonical transformations defining the $sp_2(\mbox{\ee R})$ sub algebra. 
In this case the generators are given by the classical Hamiltonian vector 
fields generated by quadratic Hamiltonians and the classical and quantum 
generators are identical [see (\ref{weyl.25}) below]. 

The classical Lie approach   
is sometimes referred to as the Poisson-Lie theory (PLT). This terminology 
is quite appropriate for generalizations as the algebraic representations 
that are relevant are written by using the Poisson and Lie product rules. 
For a similar reason we suggest that the quantum approach be referred   
to as the Moyal-Lie theory (MLT). 

In section 1 we start 
with a brief textbook discussion of the PLT. 
The Lie algebraic treatment of the quantum transformation theory starts in 
section 2 with a discussion 
of Weyl correspondence (section 2.1) and the $\star$-product. The main 
results of this paper, the MLT and the $\star$-covariance are discussed at 
length in section 2.2. The associativity property is discussed comparatively 
with respect to Poisson-Lie and the Moyal-Lie algebras in section 3.   
In the light of the classical covariance versus $\star$-covariance of the 
phase space trajectories under canonical transformations 
the time evolution deserves a specific attention. The time evolution in the MLT and 
its extended $\star$-covariance property is examined in section 4. 
Finally, we discuss the equivalence of the MLT to the standard $\star$-quantization 
through the $\star$-exponentiation in section 5. 

\subsection{The Poisson-Lie Theory and the classical phase space}
The classical Lie generators $L_{{\cal G}_\mu}$ are given by 
\begin{equation}
L_{{\cal G}_\mu}=(\partial_z {\cal G}_\mu)^T\, J \, \partial_z~,  
%\qquad {\cal G}_\mu={\cal G}_\mu(z)
\label{lie.1}
\end{equation}
where ${\cal G}_{\mu}={\cal G}_\mu(z)$'s are in the set of phase space 
functions (generating functions) 
where $\mu=1,\dots,N$ describes the generator index, $N$ being the total 
number of generators,  
$z$ denotes the $2n$ dimensional phase space row vector 
$z=(z_1,\dots,z_{2n})=(q_1,\dots,q_n; p_1, \dots, p_n)$, $\partial_z=
(\partial_{z_1},\dots,\partial_{z_{2n}})$, $T$ is the  
transpose of a row vector, and $J$ is the 
$2n \times 2n$ symplectic matrix 
with elements $(J)_{i\,j}~, ~~ i,j=(1,\dots,2n)$  
\begin{equation}
(J)_{i\,j}=\cases{1 & if j=i+n \cr 
                -1 & if i=j+n \cr 
                 0 & all other cases}~. 
\label{lie.2}
\end{equation}
The classical algebra of the phase space generating functions is defined  
over the Poisson bracket (PB)
\begin{equation}
\{{\cal G}_\mu,{\cal G}_\nu\}^{(P)}=
(\partial_z \, {\cal G}_\mu )^{T} \, J \, (\partial_z \, {\cal G}_\nu )
\label{lie.4}
\end{equation}
and that of the generators $L_{{\cal G}_\mu}$ over the Lie Bracket (LB) 
\begin{equation}
[L_{{\cal G}_\mu},L_{{\cal G}_\nu}]\equiv 
L_{{\cal G}_\mu}~ L_{{\cal G}_\nu} - 
L_{{\cal G}_\nu} ~ L_{{\cal G}_\mu}= 
L_{\{{\cal G}_\mu,{\cal G}_\nu\}^{(P)}}
\label{lie.3}
\end{equation}
where the LB is defined by the first equality.  

A classical canonical transformation (CCT) 
is considered to be a symplectic phase space map 
${\cal M}_{\vec \epsilon}: z \mapsto Z_{\vec \epsilon}(z)$ with  
$Z=(Q_1,\dots, Q_n; P_1, \dots, P_n)$ describing the new set of canonical 
phase space pairs. The symplectic condition is the invariance of the PB 
given by   
\begin{equation}
\{z_j,z_k\}^{(P)}=\{Z_j,Z_k\}^{(P)}=J_{j\,k}
\label{sympl.1}
\end{equation}
which implies that ${\cal M}_{\vec \epsilon}$ is invertible. If 
the symplectic map is continuously connected to the identity within a  
domain ${\vec \epsilon}={\vec 0}$, i.e.  
${\cal M}_{{\vec \epsilon}={\vec 0}}=\mbox{\ee I}$, Lie's 
first theorem\cite{Santilli} states that the first order infinitesimal  
change in $z_j$ is  
\begin{equation}
\delta_j(z,\delta\epsilon^{(1)})=\delta\epsilon_\mu\,L_{{\cal G}_\mu}\,z_j~. 
\label{sympl.2}
\end{equation}
where a summation over the generating function index $\mu$ is assumed. 
Finite canonical transformations can be obtained by infinitely iterating 
Eq.\,(\ref{sympl.2}) to an exponential operator as 
\begin{equation}
Z_j=\exp\{\epsilon_{\mu}\,L_{{\cal G}_{\mu}}\}\,z_j 
\label{sympl.3}
\end{equation} 
and following $[L_{{\cal G}_{\mu_k}},\,f ]=\{{\cal G}_{\mu_k},f\}^{(P)}$   
and considering $f=z_j$, 
\begin{eqnarray}
Z_j&=&z_j+\epsilon_\mu \, \{{\cal G}_\mu,z_j\}^{(P)}+{1 \over 2!}\,
\epsilon_{\mu_1} \, \epsilon_{\mu_2} \, 
\{{\cal G}_{\mu_1},\{{\cal G}_{\mu_2},z_j\}^{(P)}\}^{(P)}+\dots \nonumber \\  
&+&{1 \over k!}\,\epsilon_{\mu_1} \dots \epsilon_{\mu_k}\, 
\{{\cal G}_{\mu_1},\{\dots,\{{\cal G}_{\mu_k},z_j\}^{(P)},\dots \}^{(P)}+ 
\dots 
\label{lie.7}
\end{eqnarray} 
where summations over the repeated indices are assumed.  
Let us consider the case $N=1$. This allows us to simplify the notation and 
drop the index $\mu$. In fact, the generalization for finite $N$ is 
nontrivial because of the possibility of additional structures in the Poisson 
algebra, for instance, PB of the generators ${\cal G}_{\mu}$, where 
$\mu=1,\dots,N$,  
may be closed under a specific  algebra with $N$ generators. 
For what we present here such algebraic generalizations 
will not be needed. It can be shown that,   
if $f(z)$ has a convergent Taylor expansion in powers of  
$z$ in some domain $z \in {\cal D}$, the transformation of $f$ is  
\begin{equation}
f(Z)=\exp\{{\epsilon} \, L_{\cal G}\} \, f(z) \,
\exp\{-{\epsilon} \, L_{\cal G}\}
\label{lie.9c}
\end{equation}
where it is to be noted that Eq.\,(\ref{lie.9c}) is an operator 
relation, i.e. 
it can be multiplied on the left and/or right by any arbitrary 
function $g(z)$ without changing its validity. The transformed 
variable $Z$ and $f(Z)$ must also be well defined in the same domain 
${\cal D}$. From Eq.\,(\ref{lie.9c}) it 
follows the remarkable manifest covariance  
property\cite{Deprit,Dragt1} 
\begin{equation}
[\exp\{{\epsilon} \, L_{\cal G}\} \, f](z)\equiv 
f_{\epsilon}(z)=f([\exp\{\epsilon L_{\cal G}\}\, z])= f(Z(z,\epsilon))~, 
\label{lie.8}
\end{equation} 
and the generalized Leibniz rule 
\begin{equation}
\exp\{{\epsilon} \, L_{\cal G}\} \,[ f(z) ~ g(z) ]=
[\exp\{{\epsilon} \, L_{\cal G}\} \, f] ~ 
[\exp\{{\epsilon} \, L_{\cal G}\} \, g]=f(Z)~g(Z) 
\label{lie.9}
\end{equation}
which can be summarized in 
\begin{equation}
\exp\{{\epsilon} \, L_{\cal G}\} \,[ f(z) ~ \dots ]= 
[\exp\{{\epsilon} \, L_{\cal G}\} \, f(z) ]\,
[\exp\{{\epsilon} \, L_{\cal G}\} \,\dots]=
f(Z)\, [\exp\{{\epsilon} \, L_{\cal G}\} \dots]~. 
\label{lie.9b}
\end{equation} 
In Eq's\,(\ref{lie.8}-\ref{lie.9}) the square brackets indicate that 
the operator on the left acts on the object within the brackets only.  
In (\ref{lie.9b}) the dots indicate the possibility
of further functions to be acted upon. 
Although Eq's\,(\ref{lie.8}-\ref{lie.9b}) are different ways of writing 
Eq.\,(\ref{lie.9c}) we will keep them explicitly for that they facilitate 
a comparison with their quantum analogs.  

The MLT is a generalization of the Poisson-Lie theory  
to the non commutative phase space. 
It will be shown that, the Moyal-Lie generators can be uniquely   
derived in the quantum case with similar covariance properties when the 
role of the ordinary functional product is taken by the $\star$-product. 
As it turns out, the manifest covariance property in 
Eq.\,(\ref{lie.8}) is modified\cite{TH1} because of the non commutativity 
of the $\star$-product. 
As opposed to the standard covariance in (\ref{lie.8}) under ordinary 
multiplication   
the new covariance rule is manifest under the $\star$-multiplication of 
the functions of the phase space. The classical covariance is then 
recovered 
in the $\hbar ~~\to ~~0$ limit of the $\star$-covariance. 

\subsection{Quantum Lie generators} 
The quantum mechanical results presented here are corollaries of  
a recent work of one the present authors\cite{TH1}   
on the extended covariance in non commutative phase space 
under canonical transformations\cite{mention1}. 
Our formulation here is based on Weyl correspondence\cite{Weyl,vNWGM,Flato}
as a consistent and analytic principle between the classical and the quantum 
phase spaces. 
 
\subsubsection{The Weyl correspondence}
Let $\hat{z}=(\hat{z}_1,\dots,\hat{z}_{2n})=(\hat{q}_1,\dots,\hat{q}_n; 
\hat{p}_1,\dots,\hat{p}_n)$ describe the phase space operators satisfying,  
as usual, 
$[\hat{z}_i,\hat{z}_j]=i\hbar \, J_{i\,j}$ and zero for all other 
Lie Brackets (commutators). We adopt the same definitions for $z$ and 
$\partial_z$ as in the classical case. The Weyl map is a correspondence rule 
between the operators described by $\hat{\cal F}={\cal F}(\hat{z})$ and  
the functions $f(z)$ denoted by $\hat{\cal F} ~~\Leftrightarrow ~~ f$ and 
expressed as  
\begin{equation}
\hat{\cal F}=\int\,d\mu(z)\,f(z)\,\hat{\Delta}(z)~,\qquad 
f(z)=Tr \,\{\hat{\cal F}\,\hat{\Delta}(z)\} \label{weyl.0a} 
\end{equation}
where $d\mu(z)=\prod_{k=1}^{n} \, [dq_k \, dp_k/(2\pi \hbar)]$. 
We adopt the same definitions for $z$ and $\partial_z$ as in the 
classical case. Here 
$\hat{\Delta}(z)$ describes a mixed basis for operators and functions in 
the phase space given by      
\begin{equation}
\hat{\Delta}(z)=\int\,d\mu(\omega)\,e^{-i\omega^T\,J\,z/\hbar} \, 
e^{i\omega^T\,J\,\hat{z}/\hbar}
\label{weyl.0b}
\end{equation}
where $\omega=(\omega_1,\dots,\omega_{2n})$ is in the same 
domain as $z$, i.e. $\mbox{\ee R}^{2n}$,  
$d\mu(\omega)=\prod_{k=1}^{n}\,[d\omega_k \, d\omega_{k+n}/(2\pi \hbar)]$. 
For details on the properties of the continuous $\hat{\Delta}(z)$ basis 
we refer the reader to Ref.\,[4]. 

At a fundamental level the Weyl correspondence can be stated as 
\begin{equation}
e^{i(\omega\,J\,\hat{z})/\hbar} ~~\Leftrightarrow ~~ 
e^{i(\omega\,J\,z)/\hbar} 
\label{weyl.1}
\end{equation}
for all real $\omega$. 
 From Eq.\,(\ref{weyl.1}) the association  
for the identity, i.e. $\mbox{\ee I} 
~\leftrightarrow ~~ 1$ is unique for $\omega=0$. 
All other maps between the operators and the functions 
in the Hilbert space are derived from Eq.\,(\ref{weyl.1}) by suitable 
differential operations. Eq.\,(\ref{weyl.1}) is an explicit rule which 
leaves the algebraic properties of operators in the phase space invariant.  
Let $\hat{\cal F} ~\leftrightarrow ~~ f$ and 
$\hat{\cal G} ~\leftrightarrow ~~ g$ be associated by the Weyl 
correspondence. Then  
\begin{equation}
\hat{\cal F}\,\hat{\cal G} ~~\leftrightarrow ~~ f \star g=f\,
e^{{i\hbar \over 2}(\stackrel{\gets}{\partial_z}\,J\,
\stackrel{\to}{\partial_z}})\,g  
\label{weyl.1b}
\end{equation}
where $\star$ is a non commutative, i.e. $f \star g- g \star f \ne 0$, 
and associative, i.e. $f \star (g \star h)=(f \star g) \star h$ operation  
corresponding to the phase space symbol of the operator product.  
One way to define the $\star$ product is by its action explicitly on the 
functions of the phase space by 
\begin{equation}
f \star =f(z+{i\hbar \over 2}\,J\,\stackrel{\to}{\partial_z})~,\qquad  \quad 
\star \, f=f(z-{i\hbar \over 2}\,J\,\stackrel{\gets}{\partial_z})
\label{weyl.1c}
\end{equation}
where the arrows indicate the action of the derivatives.  
We explicitly write the correspondence for the LB as 
\begin{equation}
\bigl[\hat{\cal F},\hat{\cal G}] ~ \Leftrightarrow ~ f \star g -g \star f 
=\{f,g\}^{(M)} 
\label{weyl.2}
\end{equation}
where the superscript $M$ denotes the Moyal Bracket (MB) defined by 
the well-known expression  
\begin{equation}
\{f,g\}^{(M)}=i\hbar \, 
\{f,g\}^{(P)} +{\cal O}(\hbar^k,\partial_z^{(k)})\Bigr\vert_{3\le k}~.  
\label{weyl.3}
\end{equation}
The first term is $i\hbar$ times the PB as defined in Eq.\,(\ref{lie.4}). 
${\cal O}(\hbar^3,\partial_z^{(3)})$ represents   
higher order terms which appear in infinitely odd powers of both 
$\hbar$ and $\partial_z$.   

\subsubsection{The Unitary transformations in phase space and quantum Lie 
generators} 
All invertible transformations (including non-unitary ones) can be 
formalized in the quantum phase space\cite{TH1}. 
In this paper we will only consider 
the unitary ones. Let $\hat{\cal F}$ be an operator and $\hat{U}_{\cal A}=    
e^{i\gamma\,\hat{\cal A}}$ a one parameter unitary transformation with 
$\gamma \in \mbox{\ee R}$. In the following we prove that the unitary    
transformation of $\hat{\cal F}$ by $\hat{U}_{\cal A}$ 
in the operator space has a unique representation in the phase space given by\cite{TH1}  
\begin{equation}
\hat{\cal F}^{\prime}=
\hat{U}_{\cal A}^{\dagger}\,\hat{\cal F} \, \hat{U}_{\cal A} 
~~\Leftrightarrow ~~ 
f^\prime(z)=e^{i\gamma\,\hat{V}_{\cal A}} \, f(z)
\label{weyl.4}
\end{equation}
where, $\hat{\cal F} ~\Leftrightarrow ~ f$, 
$\hat{\cal F}^\prime ~\Leftrightarrow ~ f^\prime$ as usual, and   
$\hat{V}_{\cal A}$ is a nonlinear and Hermitian generator 
associated with a real phase space generating function ${\cal A}(z)$ by 
$\hat{V}_{\cal A}={\cal A} \star - \star {\cal A}$. 

Since $\hat{\cal A}$ is considered to be Hermitian, it can be represented  
in a symmetrically ordered series in $\hat{z}$ as,
\begin{equation} 
\hat{\cal A}={\cal A}(\hat{z}) 
=\sum_{n,m,r}\,a_{n,m,r}\,\hat{p}^n \hat{q}^m \hat{p}^r
\label{weyl.5}
\end{equation}
where $a_{n,m,r}$ are some real coefficients and we indicated the  
$\hat{p}, \hat{q}$ dependence explicitly for clarity. 
In order to prove Eq.\,(\ref{weyl.4}) we start with  
\begin{equation}
\hat{\cal F}^{\prime}=\hat{\cal F}-i\gamma \,
[\hat{\cal A},\hat{\cal F}]
+\dots +{(-i\gamma)^k \over k!}\,\underbrace{[\hat{\cal A},\dots,
[\hat{\cal A},\hat{\cal F}],\dots]}_{k}
\label{weyl.6}
\end{equation}
and calculate the commutator in the ${\cal O}(\gamma)$ term. We use 
a particular phase space representation of $\hat{z}$ 
known as the Bopp shifts and given by\cite{Bopp,Vercin,TH1} 
\begin{equation}
\hat{z}\,\hat{\Delta}(z)=\underbrace{
[z-{i\hbar \over 2}\,J\,\stackrel{\to}{\partial_z}]}_{\hat{z}_L}\,
\hat{\Delta}(z)~,\qquad
\hat{\Delta}(z)\,\hat{z}=\underbrace{
[z+{i\hbar \over 2}\,\stackrel{\to}{\partial_z}]}_{\hat{z}_R}\,
\hat{\Delta}(z) 
\label{weyl.7} 
\end{equation}
where, explicitly
\begin{equation}
\hat{z}_L={\hat{p}_L \choose \hat{q}_L}={p-{i\hbar \over 2}
\stackrel{\to}{\partial_q}  \choose 
q+{i\hbar \over 2}\stackrel{\to}{\partial_p}} ~, \qquad 
\hat{z}_R={\hat{p}_R \choose \hat{q}_R}={p+{i\hbar \over 2}
\stackrel{\to}{\partial_q}  \choose 
q-{i\hbar \over 2}\stackrel{\to}{\partial_p}}~. 
\label{weyl.7b}
\end{equation}
It is readily verified that 
\begin{equation}
[\hat{z}_{j_L},\hat{z}_{k_R}]=0~,\qquad 
[\hat{z}_{j_L},\hat{z}_{k_L}]=i\hbar \, J_{j\,k}=
[\hat{z}_{k_R},\hat{z}_{j_R}]~. 
\label{weyl.7c}
\end{equation} 
By repeatedly applying Eq's\,(\ref{weyl.7}) to the commutator of 
Eq.\,(\ref{weyl.5}) with $\hat{\Delta}$, one explicitly obtains  
\begin{eqnarray}
[\hat{\cal A},\hat{\Delta}]=\hat{V}_{\cal A}\,\hat{\Delta} 
~,\qquad \hat{V}_{\cal A}=\sum_{n,m,r}\,a_{n,m,r}\,\Bigl\{ 
\hat{p}_{L}^r \hat{q}_{L}^m \hat{p}_{L}^n -
\hat{p}_{R}^n \hat{q}_{R}^m \hat{p}_{R}^r\Bigr\}~. 
\label{weyl.8}
\end{eqnarray}
Remembering that $\hat{\cal A}$ in Eq.\,(\ref{weyl.5}) is symmetrically 
ordered, Eq.\,(\ref{weyl.8}) implies for an arbitrary phase space 
function $f(z)$  
\begin{equation}
\hat{V}_{\cal A} \, f=[{\cal A}(\hat{z}_L)-{\cal A}(\hat{z}_R)]\,f=
{\cal A}\star f-f \star {\cal A}=\{{\cal A},f\}^{(M)} 
\label{weyl.8b}
\end{equation} 
where Eq.\,(\ref{weyl.1c}) is used in the last part. 
Note that $\hat{V}_{\cal A}$ is an operator, composed of 
powers of $z, \partial_z$
 in the same way as $\hat{\cal A}$ depends on the Bopp operators 
$\hat{z}_{L,R}$, reproducing the action 
of the commutator $[\hat{\cal A},\hat{\Delta}]$ in the phase space. By 
recursive application of (\ref{weyl.8})   
the $k$'th order commutator in Eq.\,(\ref{weyl.6}) is represented by   
\begin{equation}
\underbrace{[\hat{\cal A},\dots,[\hat{\cal A},\hat{\Delta}],\dots]}_{k}=
(\hat{V}_{\cal A})^{k}\,\hat{\Delta}~,\qquad 
{\rm for~all}~ 0 \le k~. 
\label{weyl.9}
\end{equation}   
Exponentiating Eq.\,(\ref{weyl.9}) 
\begin{equation}
\hat{\cal U}^{\dagger}_{\cal A}\,\hat{\Delta}\,\hat{\cal U}_{\cal A}=
e^{-i\gamma\,\hat{V}_{\cal A}} \, \hat{\Delta}~.   
\label{weyl.10}
\end{equation}
We now use this result in the second equation in (\ref{weyl.0a}) for  
$\hat{\cal F}^\prime ~\Leftrightarrow ~f^\prime$  
\begin{equation}
f^\prime(z)=Tr\{\hat{\cal F}^\prime\,\hat{\Delta}\}=
Tr\{\hat{U}_{\cal A}^{\dagger}\, \hat{\cal F} \, 
\hat{U}_{\cal A} \hat{\Delta}\}=Tr\{\hat{\cal F} \, \hat{U}_{\cal A} \,  
\hat{\Delta} \, \hat{U}_{\cal A}^{\dagger}\}=
e^{i\gamma \, \hat{V}_{\cal A}} \, f(z)~.   
\label{weyl.11}
\end{equation}
producing the correspondence in (\ref{weyl.4}). Note that 
the unitary transformation of $\hat{\cal F}$ involves left and right 
multiplications by the transformation operator $\hat{U}_{\cal A}$ and its 
conjugate. Because of the underlying associativity these operations commute. 
The same is observed also with the exponential phase space operator 
in Eq.\,(\ref{weyl.11}). Using Eq.\,(\ref{weyl.8b}) we write 
\begin{equation}
e^{i\gamma \, \hat{V}_{\cal A}}=
e^{i\gamma \, [{\cal A}(\hat{z}_L)-{\cal A}(\hat{z}_R)]} 
=e^{i\gamma \, {\cal A}(\hat{z}_L)}\,
e^{-i\gamma \, {\cal A}(\hat{z}_R)}=e^{-i\gamma \, {\cal A}(\hat{z}_R)} \, 
e^{i\gamma \, {\cal A}(\hat{z}_L)}~. 
\label{weyl.11b1}
\end{equation}
Eq.\,(\ref{weyl.11}) is the simplest form of a canonical transformation 
acting 
on the functions of the phase space. Using Eq's\,(\ref{weyl.1c}) and 
(\ref{weyl.8b}) it can be put in the form 
\begin{eqnarray}
e^{i\gamma \, \hat{V}_{\cal A}} \, f&=&e^{i\gamma \,
[{\cal A}(\hat{z}_L)-{\cal A}(\hat{z}_R)]} \, f 
\nonumber \\
&=&
f+i\gamma \, ({\cal A} \star f -f \star {\cal A}) +{(i\gamma)^2 \over 2!}\,
({\cal A} \star {\cal A} \star f + f \star {\cal A} \star {\cal A} -
2 {\cal A} \star f \star {\cal A}) +\dots  \nonumber 
\\
&=&f+i\gamma \{{\cal A},f\}^{(M)}+{(i\gamma)^2 \over 2!} \,  
\{{\cal A},\{{\cal A},f\}\}+\dots + {(i\gamma)^k \over k!}\,
\underbrace{\{{\cal A},\dots,\{{\cal A},f\}\dots \}}_{k}+ \dots \nonumber \\
&=&e^{i\gamma \,({\cal A}\star)}\,f \, e^{-i\gamma \,(\star{\cal A})}
\label{weyl.11b}
\end{eqnarray}
which can also be deduced from (\ref{weyl.11b1}) using  
(\ref{weyl.7c}). 

There is a unique correspondence 
between $\hat{V}_{\cal A}$ and the unitary transformation 
$\hat{U}_{\cal A}$ that it represents as illustrated by 
\begin{equation}
\begin{array}{rlrlrlrl}
& f(z) ~~~~~~ & \stackrel{Weyl}{\Longleftrightarrow}& ~~~~~~~
                                                    & \hat{\cal F}
\\
e^{i\gamma\,\hat{V}_{\cal A}} 
&\Updownarrow  & ~~&    & \hat{U}_{\cal A} ~~ \Updownarrow \\  
f^\prime&=e^{i\gamma \hat{V}_{\cal A}}\,f    &
\stackrel{Weyl}{\Longleftrightarrow}  &
~~~~~~~~ &\hat{\cal F}^\prime~.
\end{array}
\label{weyl.11c}
\end{equation}
In particular, if we consider 
two such transformations acting on the phase space, 
denoting the correspondence by  
${\cal A}_1 \longleftrightarrow \hat{V}_{{\cal A}_1}$ and 
similarly for ${\cal A}_2$, we find 
\begin{equation}
\bigl[\hat{V}_{{\cal A}_1},\hat{V}_{{\cal A}_2}]=
\hat{V}_{\{{\cal A}_1,{\cal A}_2\}^{(M)}}~. 
\label{weyl.12}
\end{equation} 
Furthermore, if the operators 
$\hat{A}_k$ are the generators of an abstract Lie algebra    
the corresponding phase space operators $\hat{V}_{{\cal A}_k}$  
generate the adjoint representation in the phase space of the 
same Lie algebra. 

It is clear that $\hat{V}_{{\cal A}_k}$'s are quantum analogs of the  
classical Lie generators $L_{{\cal G}_k}$ of classical canonical 
transformations. To demonstrate this we examine the quantum analog of the 
classical covariance property in 
Eq.\,(\ref{lie.9c}-\ref{lie.8}). Since the analog of ${\cal M}_{\epsilon}$ 
is the exponentiated generator in Eq.\,(\ref{weyl.11}) we start with 
that equation. Using the correspondence in Eq.\,(\ref{weyl.1b}) we find  
\begin{equation}
f^\prime \star g^\prime=Tr\{\hat{F}^\prime\,\hat{G}^\prime \,\hat{\Delta}\}
=Tr\{\hat{U}^{\dagger}\,\hat{F}\,\hat{G}\,\hat{U}\,\hat{\Delta}\}
=(f \star g )^\prime~. 
\label{weyl.14}
\end{equation}
This equation 
is the quantum analog of (\ref{lie.9}) written more explicitly as  
\begin{equation}
[e^{i\gamma \, \hat{V}_{\cal A}} f ] \star 
[e^{i\gamma \, \hat{V}_{\cal A}} g ]
=[e^{i\gamma \, \hat{V}_{\cal A}} \, f \star g]~. 
\label{weyl.15}
\end{equation} 
where the square brackets have the same role as in 
(\ref{lie.8})-(\ref{lie.9b}). 
Considering (\ref{weyl.15}) for arbitrary $g$ the analog of 
Eq.\,(\ref{lie.9b}) is found as   
\begin{equation}
[e^{i\gamma\,\hat{V}_{\cal A}} f] \star e^{i\gamma \, \hat{V}_{\cal A}} 
=e^{i\gamma \, \hat{V}_{\cal A}} f \star 
\label{weyl.16}
\end{equation}
Eq.\,(\ref{weyl.16}) is converted into  
\begin{equation}
f^\prime \star =
e^{i\gamma \, \hat{V}_{\cal A}} f \star e^{-i\gamma \, \hat{V}_{\cal A}} 
\label{weyl.17}
\end{equation}
which is the quantum analog of Eq.\,(\ref{lie.9c}). We now examine 
the covariance properties under the action of 
$e^{-i\gamma \, \hat{V}_{\cal A}}$  
in analogy with Eq.\,(\ref{lie.8}). Suppose that $f(z)$ is expanded  
in terms of binomials $p^m\,q^n$ as  
\begin{equation}
f(z)=\sum_{0 \le (m,n)} \, f_{m,n} \, p^m \, q^n
\label{weyl.18}
\end{equation}
We use in (\ref{weyl.18}) the fact that $p^{n+m}=p^n\,p^m=[p^n \star p^m]$ 
for all $m,n$ and similarly for powers of $q$. Hence $f$ in 
Eq.\,(\ref{weyl.18}) is equivalently expressed as 
\begin{equation}
f(p,q)=\sum_{0 \le (m,n)} \, f_{m,n} \, 
\underbrace{[p\star \dots \star p]}_{m} \, 
\underbrace{[q\star \dots \star q]}_{n}~. 
\label{weyl.19}
\end{equation}
Note that, in (\ref{weyl.19}), 
the ordinary product is neither commutative nor associative 
with the $\star$-product. It can now be checked by 
using Eq's\,(\ref{weyl.11}) and (\ref{weyl.17}) that  
classical covariance described by Eq.\,(\ref{lie.8}) is 
violated
\begin{equation} 
f^\prime=[e^{i\gamma \, \hat{V}_{\cal A}} f](z) \ne f(Z)  
\label{weyl.20}
\end{equation}
where $Z=Z(z)$ are the new canonical coordinates given by 
\begin{equation}
Z=[^{i\gamma \, \hat{V}_{\cal A}} z]~. 
\label{weyl.21}
\end{equation}
We now trace back to the origin of the violation of the classical covariance  
in Eq.\,(\ref{weyl.20}). To illustrate the point we adopt two distinct 
and simple cases. As the first example let us consider $f=p \, q$. 
Our goal is to look for what causes the effect $f^\prime \ne P Q$.   
By using the properties of the $\star$-product we first write  
$f=p \star_{q,p} q+i\hbar/2$. We now apply the transformation in  
Eq.\,(\ref{weyl.11}). By Eq's\,(\ref{weyl.14}-\ref{weyl.17}) we find 
$f^\prime=P \star_{q,p} Q + i\hbar/2$. The only way in which $p q ~\mapsto ~ 
P Q$ under all such transformations as in (\ref{weyl.21}) 
is when $\star_{q,p}=\star_{Q,P}$. If this was true the covariance 
would have been manifest, since we would have had     
$P \star_{q,p} Q=P \star_{Q,P} Q=PQ-i\hbar/2$. Then inserting this in 
$f^\prime$, we would have found $f^\prime=PQ$. As the second example   
we consider $f(p,q)=p^2=[p \star p]$. Repeating the same calculation  
here we find that $f^\prime=P \star_{q,p} P \ne P^2$. 
The equality, which indicates covariance, would have been obtained for  
all transformations $Q=Q(z)$ and $P=P(z)$ if and only if 
$\star_{q,p}=\star_{Q,P}$.     

These two typical examples demonstrate that the very intrinsic property  
of the $\star$ operation, non invariance under a general canonical 
transformation, i.e. $\star_{q,p} \ne \star_{Q,P}$,  
is responsible for the loss of classical covariance in the quantum case. 
One exceptional case in which $\star_{q,p}=\star_{Q,P}$ can be identified  
corresponds to the group of linear canonical transformations, 
i.e. $Sp_2(\mbox{\ee R})$ as given by 
\begin{equation}
{P \choose Q}=g\,{p \choose q}~,\qquad g=\pmatrix{a & b \cr c & d\cr} \in 
Sp_2(\mbox{\ee R})~.  
\label{weyl.22}
\end{equation}
By conjugation, 
\begin{equation}
{\partial_P \choose \partial_Q}=(g^T)^{-1}\,{\partial_p \choose \partial_q} 
~,\qquad (g^T)^{-1}=\pmatrix{d & -b \cr -c & a\cr}~.  
\label{weyl.23}
\end{equation}
By direct inspection of Eq.\,(\ref{weyl.23}), and (\ref{weyl.1b}), 
the invariance of the $\star$ product, i.e. $\star_{q,p}=\star_{Q,P}$, 
is guaranteed by the constant coefficients. We now demonstrate,   
for the linear canonical transformations, the manifestation of the 
classical covariance in Eq.\,(\ref{lie.8}) and for a more general case 
$f(p,q)=p^m\,q^n$. By using the associativity property of $\star_{q,p}$ 
we note that $p^m \, q^n=[p_{\star}^m] \, [q_{\star}^n]$ where 
$[x^{k}_{\star}]=[\underbrace{x \star x \star \dots \star x}_{k}]$ with 
$\star=\star_{q,p}$. Since mixtures of the $\star$ and the ordinary 
product is not associative, one must perform them in the order which is 
denoted by the square brackets above. Namely, we first perform the 
calculations within each square bracket and then multiply them by using the 
ordinary multiplication. Denoting a generator of the     
linear canonical transformation $\Omega$ by $\hat{V}_\Omega$, we find that 
$f^\prime=P_{\star}^m \star \{e^{i\gamma\hat{V}_\Omega}\} \star^{-1} \, 
\{e^{-i\gamma\hat{V}_\Omega}\} Q_{\star}^n$ where we have 
$\star=\star_{q,p}$. Since $\star_{q,p}=\star_{Q,P}$ we have 
$\star \{e^{i\gamma\hat{V}_\Omega}\} \star^{-1} 
\{e^{-i\gamma\hat{V}_\Omega}\} =1$. Here $\star^{-1}$ is defined as the 
inverse of the star product which is well defined and corresponds to the 
complex conjugate of $\star$.  
We therefore have $f^\prime=[P_{\star_{P,Q}}^m] \, [Q_{\star_{P,Q}}^n]=
P^m \, Q^n$ for all nonnegative $m,n$. Hence, under linear canonical 
transformations we have the desired covariance   
\begin{equation}
[e^{i\gamma\,\hat{V}_{\Omega}} f](z)=f(Z)
\label{weyl.24}
\end{equation}    
and all other classical properties in Eq's\,(\ref{lie.8}-\ref{lie.9c}) 
can be easily derived for the linear canonical transformations here. 

Note that, the three quantum generators of the linear canonical 
algebra $sp_2(\mbox{\ee R})$ are given by 
\begin{eqnarray}
{\cal A}_1=&-{1 \over 4}(\hat{p}^2-\hat{q}^2)~,\quad \mapsto \quad
\hat{V}_{{\cal A}_1}=& {i\hbar \over 2}
\,(p \partial_q+q \, \partial_p) \nonumber \\
{\cal A}_2&={1 \over 4}(\hat{p}^2+\hat{q}^2) ~,\quad \mapsto \quad
\hat{V}_{{\cal A}_2}&={-i\hbar \over 2}\, 
(p \partial_q-q \, \partial_p) \label{weyl.25}
 \\
{\cal A}_3=&{1 \over 4}(\hat{p}\,\hat{q}+\hat{q}\,\hat{p}) ~,\quad 
\mapsto \quad
\hat{V}_{{\cal A}_3}=&{i\hbar \over 2}\, 
(p \partial_p - q \partial_q)~. \nonumber
\end{eqnarray}
The corresponding classical generators $L_{{\cal G}_k}$ are generated 
by the same quadratic polynomials ${\cal A}_k$ in (\ref{weyl.25}). 
It is observed that the quantum generators are identical  
to the classical generators (up to an $i\hbar$ factor), 
i.e. $\hat{V}_{{\cal A}_k}=i\hbar\,{L}_{{\cal G}_k}$. In ref.\,[4]  
three types of quantum phase space generators, i.e. the generators of 
$sp_2(\mbox{\ee R})$, the gauge and the contact transformations, have been 
studied. For all three cases exact solutions are possible and the  
quantum generating functions are identical to the classical ones.       
These three generators therefore generate the same canonical algebra  
and the group of transformations in the classical and quantum 
phase spaces. No other class of transformations have been found {\it yet}  
having this universal property between the classical and quantum 
cases. In the  
light of these possibilities, it was suggested that the   
canonical algebra is spanned solely by these three classes of 
generators\cite{Seligman} which is an important issue 
yet to be settled.  

The time evolution as a canonical transformation, can also be 
represented in terms of the MLT. Like in the classical case,  
the Hamiltonian function is the generator of time translations in the  
phase space. This will be examined in section 4.  
We now examine the question of associativity in the classical and the 
quantum phase space algebras.         

\subsection{Associativity and classical phase space}
The Poisson bracket as an algebraic extension of   
the abstract Lie bracket is not associative\cite{Santilli}. One way 
to understand this is to construct a Poisson product $\times$ 
which can be defined as  
\begin{equation}
\times={1 \over 2}\, 
\stackrel{\gets}{\partial_z} \, J \, \stackrel{\to}{\partial_z}  
\label{assoc.1}
\end{equation}
so that the Poisson bracket is represented as  
\begin{equation}
\{g,f\}^{(P)}=g \times f - f \times g~.  
\label{assoc.1b}
\end{equation}
The $\times$-product satisfies the left and right distributive and the 
scalar laws. Hence it is a linear algebra. Nevertheless it is 
non associative, i.e. $g \times (f \times h) \ne (g \times f) \times h$ 
which can be checked quite easily. 

Why is associativity so important in classical phase space? In pre quantum 
mechanical times the answer to this question would be {\it in order to understand 
the classical phase space as a group manifold on which a theory of Lie 
transformations can be established}. A number of attempts in this direction 
have been made 
in the past but the most serious considerations started   
after the advent of quantum mechanics and were inspired by it 
\cite{Santilli,addSantil}. 

From the classical point of view, the quantum mechanical 
$\star$-product can be realized as an associative non-commutative and 
$\hbar$ parameterized  
deformation of the $\times$-product in Eq.\,(\ref{assoc.1}). We may 
write the $\star$ product in terms of the Poisson product as,  
\begin{equation}
\star=\exp\{i\hbar\times\}~. 
\label{assoc.20}
\end{equation}
The associativity of the $\star$-product is clear by its analytic 
correspondence with     
the Lie bracket. It can be illustrative to demonstrate 
how associativity is manifested in the phase space directly. Using 
Eq.\,(\ref{weyl.7c}) we note that, the left and right multiplication in 
$\star$-product defined by Eq.\,(\ref{weyl.1c}) commute as 
\begin{equation} 
[f\star,\star g]=f(\hat{z}_R) \, g(\hat{z}_L) -
g(\hat{z}_L) \, f(\hat{z}_R)=0 
\label{assoc.2}
\end{equation}
for any two functions $f$ and $g$. Now consider three such functions 
$f, g, h$ as 
\begin{equation}
[f\star h] \star g=g(\hat{z}_L)\,f(\hat{z}_R)\,h=f(\hat{z}_R) \, 
g(\hat{z}_L) \,h
\label{assoc.3}
\end{equation}
where we used (\ref{assoc.2}) in the last step in (\ref{assoc.3}). The 
last equality implies that the differential operator $f(\hat{z}_R)$ acts 
on all functions on its right 
\begin{equation}
f(\hat{z}_R) \, g(\hat{z}_L) \,h=f(\hat{z}_R) \, [g(\hat{z}_L) \,h]=
f(\hat{z}_R) \, [h \star g]=f \star [h \star g] 
\label{assoc.4}
\end{equation}
Associativity of the $\star$-product is then proved by a comparison of 
Eq's\,(\ref{assoc.3}) and (\ref{assoc.4}). However in the case of 
Poisson product the classical version of (\ref{assoc.2}) is given by 
$[f\times, \times g] \ne 0$ for arbitrary $f,g$ causing 
Eq's\,(\ref{assoc.3}) and (\ref{assoc.4}) to be invalid. 

The associativity of the $\star$-product, combined with its manifest 
covariance under unitary transformations described by Eq.\,(\ref{weyl.14}) 
demonstrates that the $\star$-product is the algebraic partner in the 
quantum phase space of the 
ordinary functional product in the classical phase space\cite{Flato}. The 
coexistence of the $\star$ and ordinary products such as in 
(\ref{weyl.19}) is very commonly encountered  
  when the unitary 
transformation of a function $f(z)$ is to be calculated. It was proved  
in Eq's\,(\ref{weyl.24}) that, with the exception of $sp_2(\mbox{\ee R})$   
an arbitrary 
transformation does not preserve the classical covariance property. 
Here, we have an example of that from time evolution.

\subsection{$\star$-covariant time evolution in the phase space}   
In the standard formulations of the quantum mechanics 
the transformations can be represented by their actions on the states 
or on the operators or some consistent mixture of both. For the 
specific case of the time evolution these are known as 
the Schr\"{o}dinger, Heisenberg and interaction pictures respectively. We 
argue in the following in favor of adding to this gallery 
the {\it phase space picture}. 
One distinction of the phase space picture with respect 
to the other pictures is that it is essentially an operator    
picture. In this operator picture the quantum 
states in the Hilbert space that the phase space operators act on are    
represented by the density operator. The second distinction is 
that every admissible phase space operator (see section 2.1) is uniquely 
mapped onto an admissible phase space function and the operator product 
is uniquely mapped onto the $\star$-product. 
In this scheme, the density operator is uniquely mapped to a Wigner 
function. In the positivist point of view, 
the second distinction greatly facilitates the use of the conceptual tools 
to see the quantum dynamics as a classical dynamics on a deformed phase 
space\cite{TH1,Flato}. On the other hand, the opposite view is also possible. 
A negativist view can be based upon  
concentrating on the difficulties in the computations  
imported by the new rule of functional multiplication, i.e. $\star$-product.   
As it turns out, in the case when a transitive action of a transformation 
is involved, the classical and quantum cases significantly differ. A typical 
example is the time evolution.  

The time evolution in the phase space is a clear picture  
allowing a comparative analysis of the classical and quantum dynamics  
given by the same Hamiltonian ${\cal H}_t(z)$. In the quantum case we 
assume that ${\cal H}_t(z)$ can be mapped to a quantum mechanical 
Hamiltonian operator $\hat{\cal H}_t$ through the Weyl correspondence, i.e. 
${\cal H}_t(z) ~~\Leftrightarrow ~~ \hat{\cal H}_t$.  
An arbitrary function 
$f$ of the dynamical variables $z$ will be denoted by $f^{(c\ell)}_t(z)$ and 
$f^{(q)}_t(z)$ as the classical and quantum solutions corresponding to $f$. 
To facilitate the comparison further, we also assume that at a 
certain initial time  
$t_0=0$, $f^{(c\ell)}_{0}(z)=f^{(q)}_{0}(z)=f_0(z)$. The classical time 
evolution can be calculated as 
\begin{equation}
f_t^{(c\ell)}(z)={\cal T} \,e^{\epsilon \, \int_{0}^{t}\,dt^\prime \, 
L_{{\cal H}_{t^\prime}}}\,f_0^{(c\ell)}(z)~. 
\label{classtime}
\end{equation}
where ${\cal T}$ is the time ordering operator defined in the standard way by 
\begin{equation}
{\cal T} \,e^{\epsilon \, \int_{0}^{t}\,dt^\prime \, 
L_{{\cal H}_{t^\prime}}}=1+\epsilon\, \int_{0}^{t}\,dt^\prime \, 
L_{{\cal H}_{t^\prime}}+{1 \over 2!}\,\epsilon^2 \, 
\int_{0}^{t}\,dt^\prime \, L_{{\cal H}_{t^\prime}} \, 
\int_{0}^{t^\prime}\,dt^{\prime\prime} \, 
L_{{\cal H}_{t^{\prime\prime}}} \,\dots   
\label{classtime.2}
\end{equation}
If Eq.\,(\ref{weyl.4}) 
is considered for the quantum time evolution generated by 
$\hat{U}_{{\cal H}_t}$, then by  
Eq's\,(\ref{weyl.8})-(\ref{weyl.10}),   
the finite time evolution is represented in the phase space by 
\begin{equation}
f_t^{(q)}(z)={\cal T} \, 
e^{-{i\epsilon \over \hbar} \, \int_{0}^{t}\,dt^\prime \, 
\hat{V}_{{\cal H}_{t^\prime}}}\, f_0(z) ~,\qquad 
\hat{V}_{{\cal H}_{t^\prime}}={\cal H}_{t^\prime} \star - 
\star {\cal H}_{t^\prime}~. 
\label{ncov.1}
\end{equation}
For Hamiltonians quadratic in $z$ Eq's\,(\ref{classtime}) and (\ref{ncov.1}) 
yield identical results. For such quadratic Hamiltonians  
$L_{\cal H}=i\hbar\,\hat{V}_{\cal H}$ and hence  
the time evolution is covariant, i.e. 
$f_t(z)=f_0({\cal T} \, e^{-{i\epsilon \over \hbar} \, 
\int_{0}^{t}\,dt^\prime \, 
\hat{V}_{{\cal H}_{t^\prime}}}\, z)$.   
The difference arises when the third and higher order terms in $z$ are 
present in the Hamiltonian and/or in $f_0$. 
For example consider the typical case of the quartic oscillator 
${\cal H}={\cal H}_0+{\lambda \over 4}\, q^4$ where $H_0=(p^2+q^2)/2$. Using 
(\ref{ncov.1}) we have 
\begin{equation}
\hat{V}_{\cal H}=-i\hbar \, (p \,\partial_q-q\, \partial_p)+
i\,\hbar \,\lambda \, 
\Bigl[q^3 \,\partial_p-\Bigl({\hbar \over 2}\Bigr)^2 \, q \partial_p\Bigr]  
\label{ncovt.2}
\end{equation} 
There is no known method of calculating the exact analytic forms 
of the time dependence of the phase space trajectories generated by 
Eq.\,(\ref{ncovt.2}). The series expansion generates highly nonlinear 
polynomials in increasing power of $z$ at each order. 
A finite series of arbitrarily high orders can, in principle, be performed 
numerically for the binomials $p^m\,q^n$.  
Exact results can be achieved for the type of Hamiltonians 
${\cal H}=p^2/2+V(q)$ 
if a periodic kick is introduced in the interaction $V(q)$.   
The model we consider is the periodically kicked Hamiltonian  
\begin{equation}
{\cal H}(z)={p^2 \over 2}+\lambda\, {\cal V}(q)\,\delta_T({t})  
\label{ncov.0}
\end{equation}
where $\delta_{T}(t)$ is the periodic delta-kick function 
with the period $T$. 
The quantum case for (\ref{ncov.0}) 
is exactly solvable for arbitrary potentials $V(q)$ and the solution is 
identical to the classical case. We start with the classical one. 
The transformation induced by Eq.\,(\ref{ncov.0}) in one full step is 
given in terms of a time ordered product of two unitary Lie generators as 
\begin{eqnarray}
f_{n+1}^{(-)}(z)&=& e^{T\,L_{p^2/2}}\,e^{\lambda \,L_{\cal V}}\,
f_n^{(-)}(z) 
\nonumber \\
&=&f(e^{T\,L_{p^2/2}}\,e^{\lambda \,L_{\cal V}}\,z)=f(z_{n+1}^{(-)}) 
\label{ncov.1f}
\end{eqnarray}    
where the classical covariance condition is explicitly stated in the 
second part of the equation. The minus 
sign in the superscript indicates that the value of $f_n$ is calculated 
infinitesimally before the $n$'th kick.  
The time dependence for $z_n$ is given by the standard result  
\begin{equation}   
\tilde{p}_{n+1}^{(-)}=\tilde{p}_{n}^{(-)}+\kappa \, 
{\cal V}^\prime(q_{n+1})~,   
\qquad 
q_{n+1}=q_n -\tilde{p}_{n}^{(-)}~,
\label{ncov.2}
\end{equation}
with $\tilde{p}_n=T\,p_n$ and $\kappa=T\,\lambda$. 

The quantum analog of Eq.\,(\ref{ncov.1f}) in the operator space is the 
time ordered quantum evolution operator
\begin{equation}
\hat{U}_{\cal H}=e^{-{i \over \hbar}\,{\hat{p}^2 \over 2}} \, 
e^{-{i \over \hbar}\,\lambda \,{\cal V}(\hat{q})}~, 
\label{ncov.3}
\end{equation}
where ${\cal V}(\hat{q})$ is the operator corresponding to the potential 
${\cal V}(q)$. The unitary Lie generators are determined by 
Eq.\,(\ref{ncov.1}). Time ordering brings complications and therefore we 
explicitly derive $\hat{V}_{\cal H}$ from Eq.\,(\ref{weyl.4}) and 
Eq.\,(\ref{weyl.10}). Consider Eq.\,(\ref{weyl.11}) and ${\cal A}={\cal H}$ 
therein, where ${\cal H}$ is given by (\ref{ncov.0}). 
Note that, because of the unitarity of the time evolution and the 
cyclicity of the trace, the time 
ordering is effectively reversed for $\hat{\Delta}$ 
by one round of cyclic permutation of the 
operators in (\ref{weyl.11}). After permuting the 
$\hat{U}_{\cal H}$ one round, we have   
\begin{eqnarray}
\hat{\cal U}^{\dagger}_{\cal H}\,\hat{\Delta}\,\hat{\cal U}_{\cal H}&=& 
 e^{{i \over \hbar}\,\lambda \,{\cal V}(\hat{q})} \, 
e^{{i \over \hbar}\,{\hat{p}^2 \over 2}} \, \hat{\Delta} \, 
e^{-{i \over \hbar}\,{\hat{p}^2 \over 2}} \, 
e^{-{i \over \hbar}\,\lambda \,{\cal V}(\hat{q})}  \nonumber \\
&=& e^{{i \over \hbar}\,\hat{V}_{p^2/2}}\,
e^{{i \over \hbar}\,\lambda \,\hat{V}_{{\cal V}}}\,\hat{\Delta}
\label{ncov.4x}
\end{eqnarray}
yielding
\begin{eqnarray} 
e^{-{i \over \hbar}\,\hat{V}_{\cal H}} &=& 
e^{-{iT \over \hbar} \, ({p^2 \over 2} \star - \star {p^2 \over 2})} \, 
e^{-{i \over \hbar}\,\lambda \, 
({\cal V}(q)\star -\star {\cal V}(q))} \, \nonumber   
\\
f^{(q)}_{n+1}(z)&=&e^{-T\,p\,\partial_q} \, 
e^{-{i \lambda \over \hbar} \,
[{\cal V}(q-{i\hbar \over 2}\partial_p)-
{\cal V}(q+{i\hbar \over 2}\partial_p)]} \, f^{(q)}_{n}(z)~.  
\label{ncov.5}
\end{eqnarray}
By using $f^{(q)}_n(z)=z$ the transformation for the phase space  
coordinates is found to be  
\begin{equation}  
p_{n+1}=p_n+\lambda \, {\cal V}^\prime(q_{n+1})
~~~\qquad q_{n+1}=q_n-T\,p_{n}
\label{ncov.6}
\end{equation}
which is identical to the classical trajectory given by Eq.\,(\ref{ncov.2}) 
after redefining $p_n$ via $\tilde{p}_n=T\,p_n$. 
However we will keep the $\lambda$ and $T$ dependences 
as they appear in Eq.\,(\ref{ncov.6}) for reasons to be clarified later. 

The analogy with the classical trajectories is  
limited to Eq's\,(\ref{ncov.2}) and (\ref{ncov.6}) or to those obtained 
from Eq's\,(\ref{ncov.2}) and (\ref{ncov.6}) by linear canonical 
transformations. It was discussed in section 2.2 that such 
transformations create an   
equivalence class maintaining the covariance under time evolution. 

It is expected that the time evolution is non-covariant for a 
general function of phase space. By direct inspection it is observed that 
polynomials $P_r(z)$ of degree $r \le 2$ evolve identically in the 
classical and quantum cases. As a specific example we calculate 
the one time step quantum evolution of $f=p^3\,q$ which is of order $r=4$ in 
$z$. 
We find ${\cal O}(\hbar^2)$ difference between the non-covariant full quantum 
solution and the covariant classical part indicated by $(q)$ and $(c\ell)$ 
respectively. The difference is 
\begin{equation}
[p^3\,q]_{n+1}^{(q)}-[p^3\,q]_{n+1}^{(c\ell)}=
[p^3\,q]_{n+1}^{(q)}-p_{n+1}^3\,q_{n+1}= 
-{3 \over 2}\lambda\,\hbar^2 \, 
\partial_q^3\,V(q) \Bigl\vert_{q=q_{n+1}} 
\label{ncov.6b}
\end{equation}
where use have been made of 
$[p^3\,q]_{n+1}^{(c\ell)}=(p_{n+1})^3 \, q_{n+1}$ where the latter are 
,i.e. $p_{n+1}, q_{n+1}$ the same for the classical and the quantum cases. 
It turns out that a generic difference between the $(q)$  and $(c\ell)$ 
transformations of any function is always ${\cal O}(\hbar^2)$ or higher. 
One immediate observation in Eq.\,(\ref{ncov.6b}) is that it is no longer 
possible, even after suitable normalizations, to express  
$[p^3\,q]_{n+1}^{(q)}$ by the parameter $\kappa$. 
The reason is the broken classical covariance under the time evolution.  
The second observation in Eq.\,(\ref{ncov.6b}) is that, 
depending on the shape of the potential $V(q)$, the long time behavior 
of the classical and quantum solutions can be very different. If the 
third derivative of the 
potential is oscillatory, like in the standard map, the difference 
between the long time averages of the classical and quantum solutions  
in Eq.\,(\ref{ncov.6b}) vanishes. 
For polynomial potentials the right hand side in Eq.\,(\ref{ncov.6b}) is 
unbounded. In this case, and for sufficiently large $n$, the classical and 
the quantum solutions can differ significantly. 

Another feature of non covariance is about the time evolution of  
dynamical systems that are related to each other by nonlinear 
canonical transformations. Such transformations may help in understanding 
fundamentally different quantum behavior of the classically  
equivalent systems. As a typical case under the evolution by the 
Hamiltonian in (\ref{ncov.0}) consider the gauge transformation\cite{TH1}  
\begin{equation}
q ~\mapsto ~ Q=q+a\,p^3~,\qquad p ~ \mapsto ~ P=p  
\label{ncov.6c}
\end{equation}
where $a$ is a real parameter. 
Because of the nonlinear momentum dependence the quantum evolution of the 
new pair described by $P_n, Q_n$ differs from its classical 
partner. For (\ref{ncov.6c}) only the transformation of $Q$ is nontrivial.  
It can be calculated as   
\begin{eqnarray}
Q_{n+1}^{(q)}-Q_{n+1}^{(c\ell)}&=& a\Bigl([p^3]_{n+1}^{(q)}-
[p^3]_{n+1}^{(c\ell)} \Bigr) \nonumber \\
&=&a\Bigl([p^3]_{n+1}^{(q)}-p^3_{n+1} \Bigr)=
-\hbar^2 {a\lambda \over 4} \, 
\partial_q^3 V(q)\Bigr\vert_{q=q_{n+1}}~.   
\label{ncov.6d}
\end{eqnarray}
What (\ref{ncov.6d}) says is that not only the classical and the quantum 
solutions, but also  
the transformed and untransformed quantum solutions differ under 
Eq.\,(\ref{ncov.6c}). 

\subsection{Equivalence to the $\star$-exponential formalism}
The quantum Lie generator formalism as outlined in
section 2.2 can become an indispensable tool in the quantum phase space.
The equivalence of the Lie generators to the quantization by the
$\star$-exponential can be easily demonstrated.

Consider the unitary transformation $\hat{U}_{\cal A}=
e^{i\gamma\,\hat{V}_{\cal A}}$.
Expanding $\hat{U}_{\cal A}$ in power series of $\hat{\cal A}$ and
applying the Weyl map 
$\hat{\cal A}^r ~~\Longleftrightarrow ~~
\underbrace{{\cal A}\star \dots \star {\cal A}}_{r}$ we find that  
\begin{equation}
\hat{\cal U} ~\Longleftrightarrow ~ 
u=e^{i\gamma\,{\cal A}}_{\star}=1+i\gamma\,{\cal A}+{(i\gamma)^2 \over 2!}\,
{\cal A} \star {\cal A} +
{(i\gamma)^3 \over 3!}\,{\cal A} \star {\cal A} \star {\cal A}+ \dots ~.
\label{starexp.11}
\end{equation}
Eq.\,(\ref{starexp.11}) is the well known $\star$-exponential\cite{Flato}. 
The $\star$-exponential can be used in the transformation of $f$ as 
\begin{equation}
f^\prime(z)=u^{(-1)} \star f \star u=
e^{-i\gamma\,{\cal A}}_{\star} \star f(z) \star
e^{i\gamma\,{\cal A}}_{\star} ~,\qquad \star=\star_z\equiv \star_{q,p}
\label{starexp.2}
\end{equation}
This equation involves the $\star$-product of the $\star$-exponential
and it's direct calculation is notoriously difficult in yielding
analytically closed forms. It is already difficult to calculate the
$\star$-exponential and it is proven so at the most basic level, the harmonic
oscillator\cite{Flato}. Using the MLT approach the harmonic oscillator
solution can be derived quite effortlessly. The calculation can be 
found in Ref.\,[4]. By using (\ref{starexp.11}) and the results 
in section C.2 
$e^{i\gamma\,{\cal A}}_{\star} \star$ in (\ref{starexp.2}) can be
expressed as 
\begin{eqnarray}
e^{-i\gamma\,{\cal A}}_{\star} \star &=& e^{-i\gamma\,({\cal A}\star)}=
e^{-i\gamma\,{\cal A}(\hat{z}_R)} \nonumber \\
\star \, e^{i\gamma\,{\cal A}}_{\star} &=& e^{i\gamma\,(\star {\cal A})}
=e^{i\gamma\,{\cal A}(\hat{z}_L)}
\label{starexp.3}
\end{eqnarray}
If we further use 
Eq.\,(\ref{weyl.7c}) we observe that Eq.\,(\ref{starexp.3}) is
equivalent to $f^\prime=e^{i\gamma\,\hat{V}_{\cal A}}\,f$ as given 
by (\ref{weyl.11}). 

\subsection{Conclusion}
In this paper, we have introduced a new phase space approach to
Weyl quantization referred to as the Moyal-Lie theory. MLT is basically
a {\it corrected} Poisson-Lie theory by the manifestation of
associativity by the $\star$ product. The new approach is equipped with 
all other features
of classical Hamiltonian vector fields. An exceptional     
case is the need for a new concept of covariance. 
It is shown here that the covariance is equipped in the
non commutative phase space with a star product. This suggest
that one may still be able to define trajectories in the 
non commutative (quantum) sense. 

The new theory is formally equivalent to the $\star$-quantization of
Flato et al. and Bayen et al.\cite{Flato} and it may be more appealing from  
the physical point of view. This is partially due to its aspect 
which has to do with the algebraic similarities with the classical
Hamiltonian vector fields and the phase space Lie transformations.
The second aspect is the connection with the theory of quantum canonical 
transformations in the phase space which is reported elsewhere\cite{TH1}.

\section*{acknowledgments}
TH is grateful to C. Zachos (HEP Division, Argonne National Laboratory) 
for discussions. He also acknowledges the support and the hospitality of the 
DSAT group as well as the Physics Department of the University of Maryland.


\begin{thebibliography}{99}
\bibitem{mention1} The breaking of the classical covariance 
is not realized in an arbitrary way (see section B.2). Manifest invariance 
is realized when the ordinary product is replaced by the $\star$-product. 
Hence we consider this as an extension of the classical covariance.
\end{thebibliography}
\end{document}